\documentclass[journal,twoside,web]{ieeecolor}
\usepackage{jsen}
\usepackage{cite}
\usepackage{amsmath,amssymb,amsfonts}
\usepackage{algorithmic}
\usepackage{graphicx}
\usepackage{textcomp}
\usepackage[draft]{hyperref}
\usepackage{wrapfig}

\usepackage{color}
\newcommand{\RED}[1]{\color{red} #1}

\def\BibTeX{{\rm B\kern-.05em{\sc i\kern-.025em b}\kern-.08em
    T\kern-.1667em\lower.7ex\hbox{E}\kern-.125emX}}
\definecolor{abstractbg}{rgb}{0.89804,0.94510,0.83137}
\begin{document}
\title{Parkinson's Disease Recognition Using \\SPECT Image and Interpretable AI: A Tutorial}
\author{Theerasarn Pianpanit,
        Sermkiat Lolak,
        Phattarapong Sawangjai, 
        Thapanun Sudhawiyangkul$^*$ and Theerawit Wilaiprasitporn$^*$, \IEEEmembership{Member, IEEE}
\thanks{This work was supported by PTT Public Company Limited, The SCB Public Company Limited, Thailand Science Research and Innovation (SRI62W1501) and Office of National Higher Education Science Research and Innovation Policy
Council (C10F630057).}
\thanks{T. Pianpanit is with Department of Applied Radiation and Isotopes, Faculty of Science, Kasetsart University, Bangkok, Thailand.}
\thanks{S. Lolak is with Department of Clinical Epidemiology and Biostatistics, Faculty of Medicine Ramathibodi Hospital, Mahidol University, Bangkok, Thailand.}
\thanks{P. Sawangjai, T. Sudhawiyangkul, and T. Wilaiprasitporn are with Bio-inspired Robotics and Neural Engineering (BRAIN) Lab, School of Information Science and Technology (IST), Vidyasirimedhi Institute of Science \& Technology (VISTEC), Rayong, Thailand.}
\thanks{{\tt\small $^*$corresponding authors: thapanun.s at vistec.ac.th, theerawit.w at vistec.ac.th}}
}

\IEEEtitleabstractindextext{%
\fcolorbox{abstractbg}{abstractbg}{%
\begin{minipage}{\textwidth}%
\begin{wrapfigure}[18]{r}{2.7in}%
\includegraphics[width=2.6in]{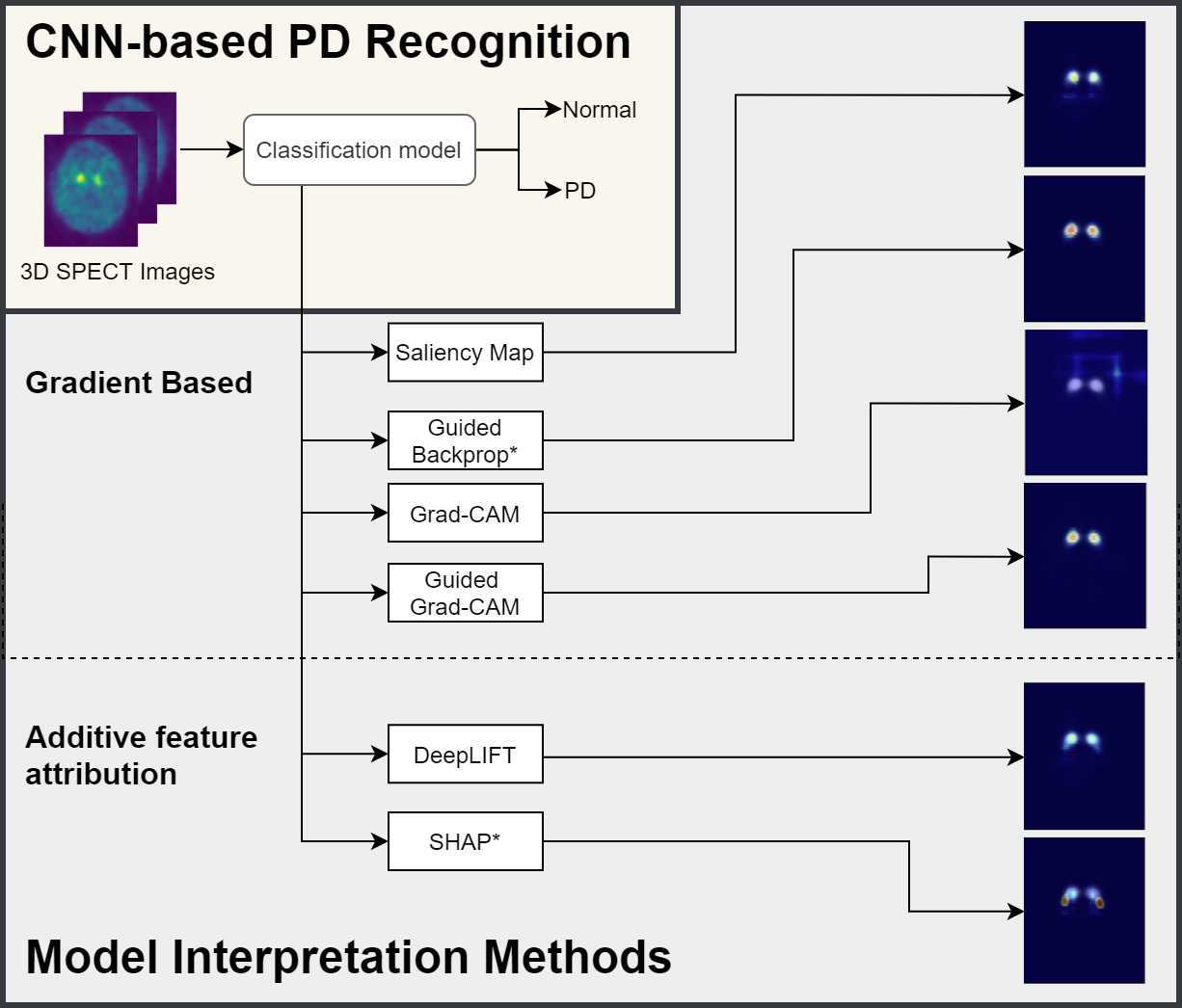}%
\end{wrapfigure}%
\begin{abstract}
\bstctlcite{bstctl:nodash}
{\RED In the past few years, there are several researches on Parkinson's disease (PD) recognition using single-photon emission computed tomography (SPECT) images with deep learning (DL) approach. However, the DL model's complexity usually results in difficult model interpretation when used in clinical. Even though there are multiple interpretation methods available for the DL model, there is no evidence of which method is suitable for PD recognition application. This tutorial aims to demonstrate the procedure to choose a suitable interpretation method for the PD recognition model. We exhibit four DCNN architectures as an example and introduce six well-known interpretation methods. Finally, we propose an evaluation method to measure the interpretation performance and a method to use the interpreted feedback for assisting in model selection. The evaluation demonstrates that the guided backpropagation and SHAP interpretation methods are suitable for PD recognition methods in different aspects. Guided backpropagation has the best ability to show fine-grained importance, which is proven by the highest Dice coefficient and lowest mean square error. On the other hand, SHAP can generate a better quality heatmap at the uptake depletion location, which outperforms other methods in discriminating the difference between PD and NC subjects. Shortly, the introduced interpretation methods can contribute to not only the PD recognition application but also to sensor data processing in an AI Era (interpretable-AI) as feedback in constructing well-suited deep learning architectures for specific applications.}
\end{abstract}

\end{minipage}}}

\maketitle

\section{Introduction}
\label{S:1}
Parkinson's disease (PD) is a chronic neurodegenerative disease caused by the nigrostriatal pathway degeneration and leads to dopamine's insufficiency in the striatum \cite{obeso10}. The characterization of the disease based on the motor symptoms are tremor, rigidity, and bradykinesia. Moreover, the non-motor symptoms which are depression, apathy, and sleep disorder, are frequently recognized. These symptoms degrade the quality of life of the people who suffer from this disease \cite{chaudhuri09}. Early and accurate diagnosis is crucial for effective treatment. The use of I123-Ioflupane SPECT or sometimes known as DaTSCAN or [123I]FP-CIT images, has become reliable as one of the PD diagnosis standards \cite{djang12}. The I123-Ioflupane has a high binding affinity for presynaptic dopamine transporters (DAT) inside the striatum. Healthy subjects are characterized by intense and symmetric uptake of the I123-Ioflupane in the caudate nucleus and putamen in both hemispheres. The striatal transaxial images should appear as the symmetric comma- or crescent-shaped. On the other hand, PD subjects are indicated by the unilateral or bilateral decrease in the uptake of the I123-Ioflupane. The striatal transaxial image often shrinks to a circular or oval shape on one or both sides. In clinical practice, diagnosis using SPECT images is usually evaluated visually and sometimes includes assistance from the semi-quantification method, which relies on computer software to acquire quantification of SPECT images \cite{badiavas11}.

The study of automated computer-aided diagnosis (CAD) of PD currently focuses on the supervised machine learning algorithm, which receives multi-dimensional input features. The machine learning methods for SPECT images classification between healthy and PD subjects from several studies show very high accuracy, generally above 90\% \cite{taylor17}. Conventional supervised machine learning for the CAD faces the difficulty of processing the images in their original form. Hand-engineering is needed in selecting the region of interest that leads to appropriate features in which the classifier can detect the patterns. Deep convolutional neural network (DCNN), which does not rely heavily on hand-engineering, has recently become a mainstream method for solving image classification problems \cite{lecun15, goodfellow16}. 

The DCNN composing the convolutional and pooling layers is inspired by the receptive fields in the visual cortex \cite{hubel62}. The resemblance of the DCNN and the primate visual stimuli processing has also been evaluated using the last convolutional layer's features from the DCNN, and the inferior temporal cortex neural responses \cite{cadieu14}. In addition, the progress in hardware, software, and algorithm parallelization, which reduces the training time to process a massive collection of multi-dimensional data, allows DCNN to become a high-performance tool in medical image recognition \cite{litjens17, duncan20}. Further investigation shows that DCNN still gives high classification accuracy even without the need for spatial normalization procedure \cite{martinez18}. However, it is still unclear which regions in the images are being detected by the model and whether the DCNN understands the pattern in the same way as the expert's visual interpretation. Unlike the conventional machine learning models in which each input feature is hand-designed and the models are decomposable into interpretable components, the complexity of the DCNN seems to diminish its interpretability. Also, the EU’s General Data Protection Regulation (GDPR), Recital 71, which gives citizens a ``right to explanation'' will make the ``black box'' approaches hardly suitable in clinical diagnosis \cite{ras18}.

Several DCNN model interpretation methods have been developed to visualize or interpret the DCNN so that the attention map can be generated to understand the essential pixels of the input image. These methods were used to interpret the model's decision and increase the credibility of the DCNN diagnosis results in several types of medical image \cite{martinez18, esteva17, lee19}. However, due to the variety of model interpretation methods, there is no evidence of which methods can provide the most reliable interpretation results for medical image applications.

{\RED This tutorial aims to demonstrate the procedure for selecting the most suitable interpretation method for SPECT image PD recognition model. The contributions of this tutorial are as follows:
\begin{enumerate}
\item We provide an overview on the recent PD recognition model, and provide a step-by-step approach to implement four DCNN models.
%we provide a brief introduction to the traditional method and DCNN models for PD recognition. Next, we give an example scenario by training four DCNN models and comparing the classification performance on the PD diagnosis. Note that this tutorial will focus on the interpretation method and will not go deeply into PD classification models. However, this part could give a general idea to those who are new or not actively involved in PD recognition.
\item We incorporate six well-known interpretation methods to four DCNN models, display each method's visual interpretation result, and demonstrate the methods for evaluating the interpretation performance. To the best of our knowledge, this is the first attempt to explore the interpretation methods that is suitable for using with SPECT image PD recognition model.
\item We propose a method to utilize the interpreted feedback to aid in model selection.
\end{enumerate}
}

The code for all four DCNN models with six interpretation methods was uploaded and can be download publicly\footnote{https://github.com/IoBT-VISTEC/PPMI\_DL, We will publish all source codes and data sources immediately after getting an acceptance letter from SJ}. Furthermore, the introduced deep neural network interpretation methods can contribute to the future of data processing in an AI Era (interpretable-AI) as one of the core modules in sensors-related studies. For example, Grezmak \textit{et al.} had reported the interpretable CNN for a machine fault diagnosis \cite{machine_lrp}, Alharthi \textit{et al.} had reported an interpretable time series model for gait-induced ground reaction force (GRF) in Parkinson’s disease (PD) recognition \cite{gait_lrp}, and Lee \textit{et al.} had utilized interpretable AI in glucose management for diabetes patient \cite{9115809}. All of the examples demonstrate the usefulness of the model interpretation methods as feedback in constructing well-suited deep learning architectures.

\section{PD Recognition Methods and an Example Scenario}
\label{S:2}

\subsection{Traditional Classification Method}
The most commonly used features for the traditional classification method are the striatal binding ratios (SBR) from both left and right caudate and putamen, which relate to the ratio of the target region and the reference region. These features were classified with the probabilistic neural network, decision tree \cite{palumbo10}, and support vector machine (SVM) \cite{palumbo14, prashanth14}. Other new methods have been developed to find the features from region of interest (ROI), including shape analysis and surface fitting \cite{prashanth17}, mean ellipsoid uptake and dysmorphic index \cite{augimeri16}, Haralick texture features \cite{martinez13}, principal component analysis (PCA) \cite{towey11}, independent component analysis (ICA) \cite{martinez14}, partial least squares decomposition \cite{segovia12}, empirical mode decomposition with PCA or ICA \cite{rojas13} or circularity features obtained from DAT \cite{shiiba20}. These new types of features seem to give the best accuracy with the SVM classifier. Furthermore, the image voxels within the ROI are also used directly as the input features with SVM \cite{illan12, oliveira15}, logistic lasso \cite{tagare17}, and single-layer neural network \cite{zhang17} classifiers.

In this tutorial, we utilize the most commonly used SBR feature with an SVM classifier as an example of the traditional classification method. The SBR \cite{sbr13} can be calculated by first applying the standard Gaussian 3D 6.0 mm filter to the final preprocessed images. These images were then normalized to standard Montreal Neurologic Institute (MNI) space so that all scans are in the same anatomical alignment, followed by identifying the transaxial slice with the highest striatal uptake. Then, the 8 hottest striatal slices around it were averaged to generate a single slice image. Regions of interest (ROI) were then selected for left and right caudate, and left and right putamen. The occipital cortex was selected as the reference region. Count densities for each region were extracted, and SBR is calculated as

\begin{equation}
\textrm{SBR of target region} = \frac{\textrm{Target region count density}}{\textrm{Reference region count density}}-1.
\end{equation} 

The SBR of each subject can be obtained from the PPMI database alongside the SPECT images. It was proved that applying SBR to SVM gives very high accuracy \cite{prashanth14}; therefore, we will use this classification method as a baseline for comparing and evaluating with the deep learning approach.
 
\begin{table*}
	\caption{\RED Summarize of traditional classification method (upper) and deep learning (lower) for Parkinson's disease SPECT image classification.}
	\centering
\resizebox{0.99\textwidth}{!}{%
	\begin{tabular}{ l c l l l l }
     Reference &	Number of subjects 	&	Feature & Classifier & Dataset & Accuracy \\
     & PD:Control & & & & \\
	\hline
    Palumbo et al. 2010 \cite{palumbo10} & 127 : 89 & Striatal binding ratios & PNN, CT & Private dataset & PNN: $96.6 \%$, CT: 93.5  \% \\
    Towey et al. 2011 \cite{towey11} & 79 : 37 & PCA decomposition of striatal region & Naive-Bayes & Private dataset & 94.8\%   \\
    Oliveira et al. 2015 \cite{oliveira2015computer} & 445 : 209 & Voxel-base-feature & SVM & PPMI &  97.86\%   \\
    
    Oliveira et al. 2018 \cite{oliveira2018extraction} & 443 : 209 & SBR, CBP, PBP, SBP, PCR, LSR, VSR & SVM, kNN, LR & PPMI &  SVM: 97.90\%, kNN: 97.20\%, LR: 96.90\%   \\
    
    \hline
    \\
    
     Reference &	Number of subjects 	&	Feature & Classifier & Dataset & Accuracy \\
     & PD:Control & & & & \\
	\hline    
    Martinez et al. 2017 \cite{martinez2017} & 158 : 111 & None & DCNN & PPMI &  95.50\%   \\
    
    Choi et al. 2017 \cite{choi17} & 431 : 193 (PPMI) & None & PDNet & PPMI &  96.00\%   \\
    
            & 72 : 10 (SNUH) & None & PDNet & SNUH (Private dataset) &  98.8\%   \\

    Wenzel et al. 2019 \cite{wenzel2019automatic} & 438 : 207 & None & DCNN & PPMI &  97.20\%   \\

    Ortiz et al. 2019 \cite{ortiz2019parkinson} & 158 : 111 & Isosurface & LeNet, AlexNet & PPMI &  LeNet: 95.10\%, AlexNet: 95.10\%   \\

    Mohammed et al. 2021 \cite{mohammed2021easy} & 1359 : 1364  & None & DCNN & PPMI &  99.34\%   \\
     & (1023 of control images were augmented images) &  &  &  &     \\

	\hline
    \multicolumn{6}{l}{SBR: Striatal binding ratios, CBP: Caudate binding potential, PBP: Putamen binding potential, SBP: Striatal binding potential, PCR: Putamen-to-caudate ratio, LSR: Length of the striatal region, 
    %VSR: volume of the striatal region, kNN: k-Nearest neighbor, LR: Logistic regression, PNN: Probablistic neural network, CT: Classification tree" 
    }\\
    \multicolumn{6}{l}{
    %SBR: Striatal binding ratios, CBP: Caudate binding potential, PBP: Putamen binding potential, SBP: Striatal binding potential, PCR: Putamen-to-caudate ratio, LSR: Length of the striatal region, 
    VSR: Volume of the striatal region, kNN: k-Nearest neighbor, LR: Logistic regression, PNN: Probablistic neural network, CT: Classification tree }\\
	\end{tabular}}
	\label{tbl-ref01}
\end{table*}

%%%%%%%%%%%%%%%%%%%%%%%%%%%%%%%%%%%%%%%%%%%%%%%%%%%%
\begin{figure}
\centering
\includegraphics[width=0.46\textwidth]{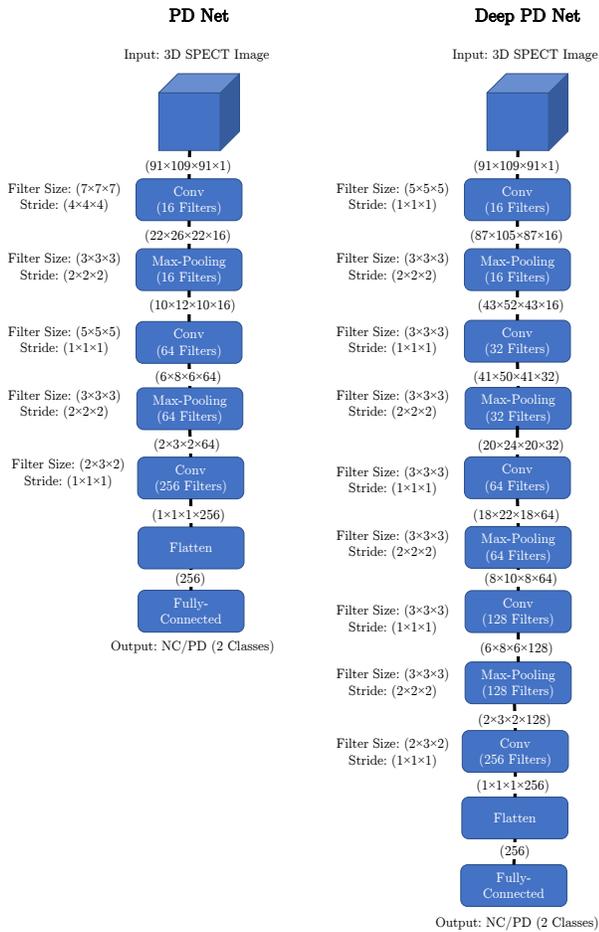}
\caption{Structure of PD-Net and Deep PD Net used as examples in this tutorial with the details of the size and number of convolution and max-pooling filters. The PD Net has been modified in the last convolution layer so that the image from the database can be used directly without the need for zero-padding.}
\label{fig:DCNN}
\end{figure}
%%%%%%%%%%%%%%%%%%%%%%%%%%%%%%%%%%%%%%%%%%%%%%%%%%%%

\subsection{CNN Architectures for PD recognition}

There are several DCNN based models for PD recognition using SPECT images. From 2017, Martinez-Murcia \textit{et al.} proposed utilization of DCNN on SPECT image to diagnose PD \cite{martinez2017}. They trained their model with 301 SPECT images (158 PD, 111 normal control (NC), and 32 scans without evidence for dopaminergic deficit (SWEDD)) from the PPMI database, and their network achieved up to 95.5\% accuracy (96.2\% sensitivity). Choi \textit{et al.} proposed a deep DCNN model ``PD Net'' which was trained with the whole volume of SPECT images to discriminated the PD subjects from NC subjects \cite{choi17}. The model was trained with 624 subjects (431 PD and 193 NC) from the PPMI database, resulting in 96.0\% accuracy (94.2\% sensitivity) comparable to the evaluation from the experts. 

Later in 2018, Wenzel \textit{et al.} proposed a large DCNN model with 2,872,642 parameters trained by 645 subjects from PPMI (438 PD and 207 NC). Despite the fact that this model yield 97.7\% accuracy (96.6\% sensitivity), slightly better than PD Net, the model is large and resource-consuming. Recently in 2021, Mohammed \textit{et al.} proposed the present state-of-the-art model with minimal architecture \cite{mohammed2021easy}. Their model consisted of three convolutional layers with a filter size of $(3\times3)$ and two dense layers. Their model's input image was normalized to enhance the ROI and provide a distinguishing feature to the model. A 10-fold cross-validation was used to evaluate the performance of the model. This state-of-the-art model was trained by 2723 SPECT images from the PPMI database (1359 PD and 1364 NC) and can provide 99.3\% accuracy (99.0\% sensitivity). 
 
\subsection{Model implementation: Example Scenario} 
{\RED
For a demonstration, we incorporate four different DCNN architectures based on PD Net \cite{choi17} for comparing in both classification and interpretation performance. As a tutorial, we choose these four DCNN architectures so that the classification performance is not significantly different and difficult to evaluate. Later in the tutorial, we will show another benefit of model interpretation: to interpret feedback as an evaluation metric.
%Noted that these four DCNN were chosen as only examples in this tutorial.
For further study and development, we suggest utilizing the state-of-the-art model \cite{mohammed2021easy}.

\subsubsection{Models description}
The first model in this tutorial is the PD Net illustrated on the left-hand side of \autoref{fig:DCNN}. In the original PD Net, zero-padding was applied to make the image's size equal in all dimensions. However, this tutorial does not include zero-padding so that the images are all in their original form. Thus, a slight modification of the filter size is made in our PD Net model. PD Net model is composed of three 3D convolution layers connected with a single fully connected layer. Each 3D convolution layer has a different setup of filter size and stride, but all 3D convolution layers have Rectified Linear Unit (ReLU) activation layer and a max-pooling layer with $(3\times3\times3)$ pool size and stride of 2 attached. The first 3D convolution layer has 16 filters with a size of $(7\times7\times7)$ and a stride of 4. After the first pooling, images are fed to the second 3D convolution layer, which has 64 filters with a size of $(5\times5\times5)$ and a stride of 1. Finally, a 3D convolution layer with 256 filters of size $(2\times3\times2)$ and a stride of 1 is attached. This layer produces 256 features, which then fully-connect to 2 output nodes to discriminate the extracted features.  

The second model is a modified PD Net architecture by increasing the network depth as shown on the right-hand side of \autoref{fig:DCNN}. We refer this model as ``Deep PD Net''. In this model, the filter size of both the 3D convolution and max-pooling layers was designed so that the last layer before the fully-connect layer gives 256 features, the same as PD Net. 

The third and fourth models are PD Net and Deep PD Net with batch normalization. The batch normalization layer was added to follow each ReLU layer. Batch normalization was proposed to accelerate DCNN's training and was first applied with the image classification task \cite{ioffe15}. It can achieve the same accuracy with a much lower learning rate; thus, it reduces the number of epochs for training.  
%The training parameters for all DCNN models are elaborated in \nameref{A:1}.

\subsubsection{SPECT image dataset}
The public SPECT image dataset commonly used in PD recognition studies \cite{Klyuzhin29,ortiz19,wenzel2019automatic,mohammed2021easy,choi17} are from Parkinson’s Progression Markers Initiative (PPMI) database \cite{ppmi12}. PPMI is a study from the collaboration of research centers designed to identify PD progression biomarkers and to provide essential tools to improve PD therapeutics. All SPECT scan data acquired from every center undergo the same preprocessing procedure before they are publicly shared via the database \cite{sbr13}. SPECT raw projection data was imported to a HERMES\footnote{Hermes Medical, Stockholm, Sweden} system for iterative reconstruction using the HOSEM software. Iterative reconstruction was done without applying any filter. The HOSEM reconstructed files were then transferred to PMOD\footnote{PMOD Technologies, Zurich, Switzerland} for further processing. Attenuation correction ellipses were drawn on the images and a Chang 0 attenuation correction was applied. The final 3D-volume SPECT image with the voxel size of $2\times 2\times 2 \;\mathrm{mm^3}$ and the dimension of $91\times 109\times 91$ can be directly downloaded from the publicly shared PPMI database.
}
\subsubsection{Data selection and pre-processing}
{\RED A total of 607 subjects with clinical characteristics summarized in \autoref{tbl-para} were selected for training the models in this tutorial. Since PPMI is the longitudinal study of the PD subject, only the earliest SPECT image was selected for each subject and we selected one SPECT image per subject. The selected data has more PD class than NC class, since the PPMI database provide more data from PD class than NC class. This make the data imbalance, which in some work, the data augmentation on NC class is utilized to balance the data \cite{mohammed2021easy}. In this tutorial, we will not cover on the data augmentation method. After obtaining SPECT images from PPMI, the min-max normalization in the range $[0,1]$ is applied.}

\subsubsection{Training and testing process}
{\RED All the DCNN models were implemented with Keras \cite{chollet15}, an open-source deep learning library written in Python and running on top of Tensorflow \cite{abadi17}. The models were trained for 30 epochs using Stochastic Gradient Descent. The momentum parameter was set to 0.9. The learning rate was initially $1 \times 10^{-4}$ and logarithmically decreased to have $1 \times 10^{-6}$ at the final epoch. Additionally, weight parameters in the model were initiated with a Glorot initialization \cite{glorot10}. The loss function also is weighted for class imbalance during the training. These training parameters are the same with \cite{choi17} and every model uses the same parameters for a fair comparison.}

The data were divided into training, validation, and testing set with a ratio of 80:10:10. During the training, the model uses the validation set to tune the model to reach the best classification performance. The experiment is carried out using 10-fold cross-validation. The best model that the validation set provides in each fold is used to calculate both classification and interpretation performance by applying it to the testing set.

%%%%%%%%%%%%%%%%%%%%%%%%%%%%%%%
\begin{table}
	\caption{Clinical details of all subjects used in this tutorial.}
	\centering
\resizebox{0.45\textwidth}{!}{%
	\begin{tabular}{ l l l }
     &	Parkinson's disease	&	Normal Control \\
     &	(n=448)	&	(n=159) \rule[-0.9ex]{0pt}{0pt}\\
	\hline
    Age				&	61.6 $\pm$ 9.8	& 	60.5 $\pm$ 11.3	\rule{0pt}{2.6ex}\\
    Sex (M/F)		&	288/160	& 	112/47  \\
    MDS-UPDRS part III  & 21.3 $\pm$ 9.5 &   \\
    Hoehn and Yahr stage & 1.6 $\pm$ 0.5  &   \rule[-0.9ex]{0pt}{0pt}\\
	\hline
	\end{tabular}}
	\label{tbl-para}
\end{table}

The classification performance of each model is reported using the 10-fold cross-validation. In addition to the accuracy, sensitivity and specificity are used as metrics to compare each model. They are defined as
\begin{equation}
\mathrm{Sensitivity = \frac{ True\; positive}{Total\; positive}},
\end{equation}
\begin{equation}
\mathrm{Specificity = \frac{\mathrm True\; negative}{Total\; negative}}.
\end{equation}

Results that were acquired using SBR as the input feature along with the SVM classifier were used as the benchmark to compare with the deep learning method, which uses whole volume SPECT image as the input feature with DCNN as the classifier. Four types of DCNN architecture were designed based on the PD Net \cite{choi17} and all of them are described in the previous section. The mean $\pm$ STD of accuracy, sensitivity, and specificity calculated from 10-fold of a testing set, are shown in \autoref{tbl-acc}. The accuracy varies from 95\% to 96\% with the deep learning approaches, giving a slightly higher accuracy than the SVM model. Deep PD Net with batch normalization has the highest accuracy with 96.87\%. The sensitivity of each model was not significantly different. However, we can see the improvement of the specificity from 93\% to 97\% of the Deep PD Net model.

\begin{figure}
\centering
\includegraphics[width=0.46\textwidth]{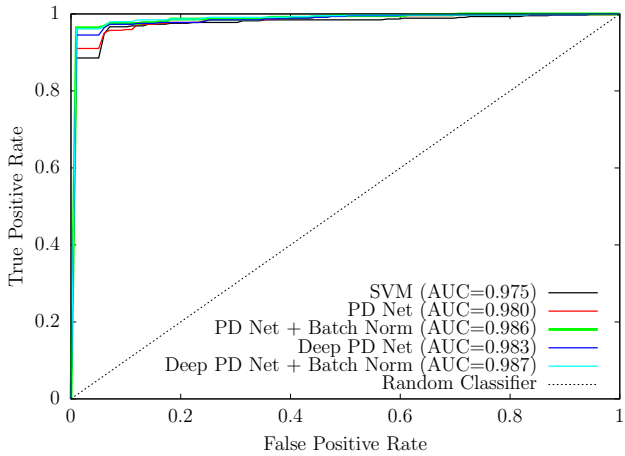}
\caption{ROC curve for each model.}
\label{fig:roc_curve}
\end{figure}

McNemar's test \cite{dietterich98} was used to compare between SVM and DCNN models, and the $p$-value from this test can not reveal any statistical difference in the classification performance. Thus, we further investigate the ROC curve as shown in \autoref{fig:roc_curve}. It reveals a trend of a higher AUC value of DCNN than that of SVM. The Deep PD Net with batch normalization has the highest AUC value, which is 0.987.

%%%%%%%%%%%%%%%%%%%%%%%%%%%%%%%
\begin{table*}
	\centering
	\caption{Classification performance of SVM, PD Net and Deep PD Net.}
\resizebox{0.8\textwidth}{!}{%
	\begin{tabular}{ l l l l l l l}
    Method		&	Input Feature	&	Accuracy &	Sensitivity	&	Specificity &	\\
	\hline
SVM & SBR Ratio & 95.55 $\pm$ 2.48 & 96.90 $\pm$ 2.61 & 92.29 $\pm$ 7.73 & \\
PD Net & SPECT & 95.39 $\pm$ 2.88 & 95.97 $\pm$ 3.30 & 93.75 $\pm$ 6.23 & \\
PD Net + Batch Norm & SPECT & 96.54 $\pm$ 2.63 & 96.88 $\pm$ 3.20 & 95.66 $\pm$ 6.09 & \\
Deep PD Net & SPECT & 96.71 $\pm$ 2.32 & 97.10 $\pm$ 2.35 & 95.42 $\pm$ 4.40 & \\
Deep PD Net + Batch Norm & SPECT & 96.87 $\pm$ 2.13 & 96.42 $\pm$ 3.01 & 97.89 $\pm$ 3.61 & \\
	\hline
	\end{tabular}
}
	\label{tbl-acc}
\end{table*}
%%%%%%%%%%%%%%%%%%%%%%%%%%%%%%%%%%%%%%%%%%%%%%%%%%%%

\section{Model Interpretation Methods}
\label{S:3}
\subsection{Interpretation Methods Overview}

Despite the fact that DCNN models can provide highly accurate classification results, due to DCNN's black-box nature, it is difficult to directly explain the importance of the input features that lead to high classification performance. Model interpretation methods have been used to reveal the feature importance and assess the trust of the model prediction results. Hence, the primary purpose of the interpretation method is to calculate the ``contribution score'' \cite{shrikumar17} of the input features. Vastly used model interpretation methods for DCNN can be categorized into two major groups. The first one is the gradient-based method, which focuses on using backpropagation to calculate the gradient that can be implied back to be the input score of the target class's input features. The other group is the additive attribution methods, which alternatively construct a simpler model to explain the complex model. Well-known current methods belonging to these two major groups are discussed below.

\subsubsection{Gradient based method}

The core concept of deep learning is to calculate the gradient of the loss function with respect to all the model's weights and biases. These gradients can be used to compute the relation between the input feature and the output prediction class. We categorize the interpretation methods that directly use these gradients from the original model as the gradient-based method.

\textbf{Direct backpropagation (Saliency map):} Backpropagation is a method to compute gradients of the loss function for all weights in the network. These gradients can also be backpropagated to the input data layer, which contributes the most to the assigned class.  This is done by computing the gradient of the output category with respect to a sample input image \cite{simonyan13}. If we define input features as $x$ and score for predicting class $c$ as $S^c$, the map of the contribution score is calculated as
\begin{equation}
L^c_\mathrm{Saliency\;map} = \frac{\partial S^c}{\partial x}
\end{equation}

\textbf{Guided backpropagation:} For the direct backpropagation, the gradient of the loss function with respect to the parameter of layer $l+1$ is used to calculate the gradient of the loss function with respect to the parameter of layer $l$. In guided backpropagation, the same calculation with the direct backpropagation is used, but if the gradient of layer $l+1$ is negative, the gradient of layer $l$ is set to zero \cite{springenberg14}. In other words, this method includes the guidance signal to the deeper layer during the backpropagation resulted in the remarkable improvement of the contribution score map.

\textbf{Grad-CAM:} Global average pooling (GAP) is the sum of all the values in a feature map at the last convolution layer. It was proposed to replace the fully-connected layers of the DCNN. GAP reduces the total model parameters and results in preventing the overfitting from the fully-connected layers in some cases. For a 2D input image, the GAP of the $k^\mathrm{th}$ feature map $A^k$ can be calculated from the sum of the 2D elements $i,j$ or can be written as
\begin{equation}
G^k  = \sum\limits_{i} \sum\limits_{j} A^k_{ij}.
\end{equation}
The score of predict class $c$ then becomes 
\begin{equation}
S^c = \sum\limits_{k } \sum\limits_{i} \sum\limits_{j} w^c_k A^k_{ij},
\end{equation}
where $w^c_k$ is the weight of $A^k_{ij}$ to predict class $c$. By examining this equation, class activation map (CAM) can be defined as 
\begin{equation}
\mathrm{CAM} =\sum\limits_{k } w^c_k A^k_{ij},
\end{equation}
which shows the 2D map of the score that predict class $c$. CAM represents the input feature's contribution score by resizing this 2D map to the original input image. It also has a remarkable ability for object localization of the predict class \cite{zhou16}. However, the structure of GAP tends to reduce the model classification performance. The Gradient-weighted Class Activation Mapping (Grad-CAM), which is a generalized form of CAM, was proposed to handle the issue \cite{selvaraju17}. Grad-CAM directly calculates the gradient using the backpropagation from each neuron of the last convolution layer feature map, which can be written as $\partial S^c / \partial A^k_{ij}$. Then, these gradients are summed within the $k^\mathrm{th}$ feature map to generate the weight of each map and predict class $c$, which can be written as;
\begin{equation}
\alpha _k^c = \sum\limits_i {\sum\limits_j {\frac{{\partial {S^c}}}{{\partial A_{ij}^k}}} }
\end{equation}
Then Grad-CAM of class $c$ can be generated from
\begin{equation}
L_{{\mathrm{Grad - CAM}}}^c = \mathrm{ReLU} \left( {\sum\limits_k {\alpha _k^c{A^k}} } \right).
\end{equation}
ReLU function is used to remove the negative contribution scores because Grad-CAM wants to consider only the input features that increase the prediction score of class $c$. Due to the direct use of the gradient from the backpropagation, Grad-CAM can be applied to interpret any type of DCNN (e.g., DCNN with recurrent neural networks) without any modifications to the DCNN model. 

\textbf{Guided Grad-CAM:} The use of the last convolution layer of the Grad-CAM can provide a more accurate location of the relevant image regions. However, this last layer does not maintain enough resolution to provide a fine-grained importance feature. Although the guided backpropagation method provides the contribution scores of every individual pixel of the input image, it lacks the localization capability. In order to get the best outcome, it is possible to fuse guided backpropagation with Grad-CAM to create Guided Grad-CAM that has both high-resolution and high capability to locate the related image area.

\subsubsection{Additive feature attribution method}

When the model becomes more complex, the original model can hardly be used to explain its results. The best way to explain the model is to generate a simpler explanation model from the original model's approximation. 
By giving $f(x)$ to be the original model, $x$ to be the original input, $g(x')$ to be the explanation model, and $x'$ to be the simplified input, the equation used to explain the original model can be written as $g(x') = f(x)$. The simplified input must be able to map to the original input through a mapping function $x=h_x(x')$. The simplest way to represent the explanation model is to let the simplified input be the binary vector, representing the presence or absence of the input features. For the image classification task, these input features can be pixels or super-pixels. This method of generating the explanation model is defined as the additive feature attribution method \cite{lundberg17, lundberg16}, in which the explanation model $g$ is written as
\begin{equation}
\label{eq:additiveFeature}
g(x') = {\phi _0} + \sum\limits_{i = 1}^M {{\phi _i}{{x'}_i}},
\end{equation}
where $x' \in \lbrace 0,1\rbrace ^M$, $M$ is the number of simplified input features, and $\phi_i \in \mathbb{R}$. This method approximates the output $f(x)$ by using $\phi_i$ which is the ``attribution'' or ``contribution score'' from each input feature. Two well-known interpretation methods which are based on the concept of \autoref{eq:additiveFeature} are discussed below.

\textbf{DeepLIFT:} Deep Learning Important FeaTures (DeepLIFT) is an interpretation method that avoids discontinuity of the gradient-based approach in approximating the feature contribution to the output \cite{shrikumar17}. By giving reference to the input and output, the contribution scores can be calculated from the difference using this reference. If $x_i$ and $f(x)$ are input feature and model output, $x_{i0}$ and $f(x_0)$ are reference input feature and reference model output, then $\Delta y = f(x)-f(x_0)$ and $\Delta x_i = x_i - x_{i0}$ are defined as the difference between the reference and model output and input feature. DeepLift assigns the attribution of $\Delta x_i$ as $C_{\Delta x_i \Delta y}$ and uses the summation of these attributions to give the value of $\Delta y$, which can be written as;
\begin{equation}
\label{eq:sum-to-delta}
\sum\limits_{i = 1}^M C_{\Delta x_i \Delta y} = \Delta y.
\end{equation}
By comparing this with \autoref{eq:additiveFeature} with $f(x_0) = \phi_0$ and $C_{\Delta x_i \Delta y} = \phi_i$, DeepLIFT can be categorized as the additive feature attribution method. DeepLIFT uses rules, that are based on the structure of deep learning network, to assign the attribution from each input feature. Thus, DeepLIFT is ``model-specific'' in the approximation of the contribution score. DeepLIFT also shown to be the modify form with better performance compare to another model-specific method called ``layer-wise relevance propagation'' \cite{bach15}.

\textbf{SHAP:} SHapley Additive exPlanation (SHAP) was designed to simplify any complex model, not restricted to any model structure \cite{lundberg17}. For SHAP, Shapley values are used for the contribution score, and they are the only set of values that satisfy the properties of the additive feature attribution or \autoref{eq:additiveFeature}. SHAP proposes a way to approximate the Shapley value by minimizing the objective function that satisfies all the properties of \autoref{eq:additiveFeature}. This objective function does not constrain any model parameters and only use the result from the model output. Thus, SHAP becomes ``model-agnostic'' in the approximation of the contribution score.

%%%%%%%%%%%%%%%%%%%%%%%%%%%%%%%%%%%%%%%%%%%%%%%%%%%%
\begin{figure*}
\centering
\includegraphics[width=0.9\textwidth]{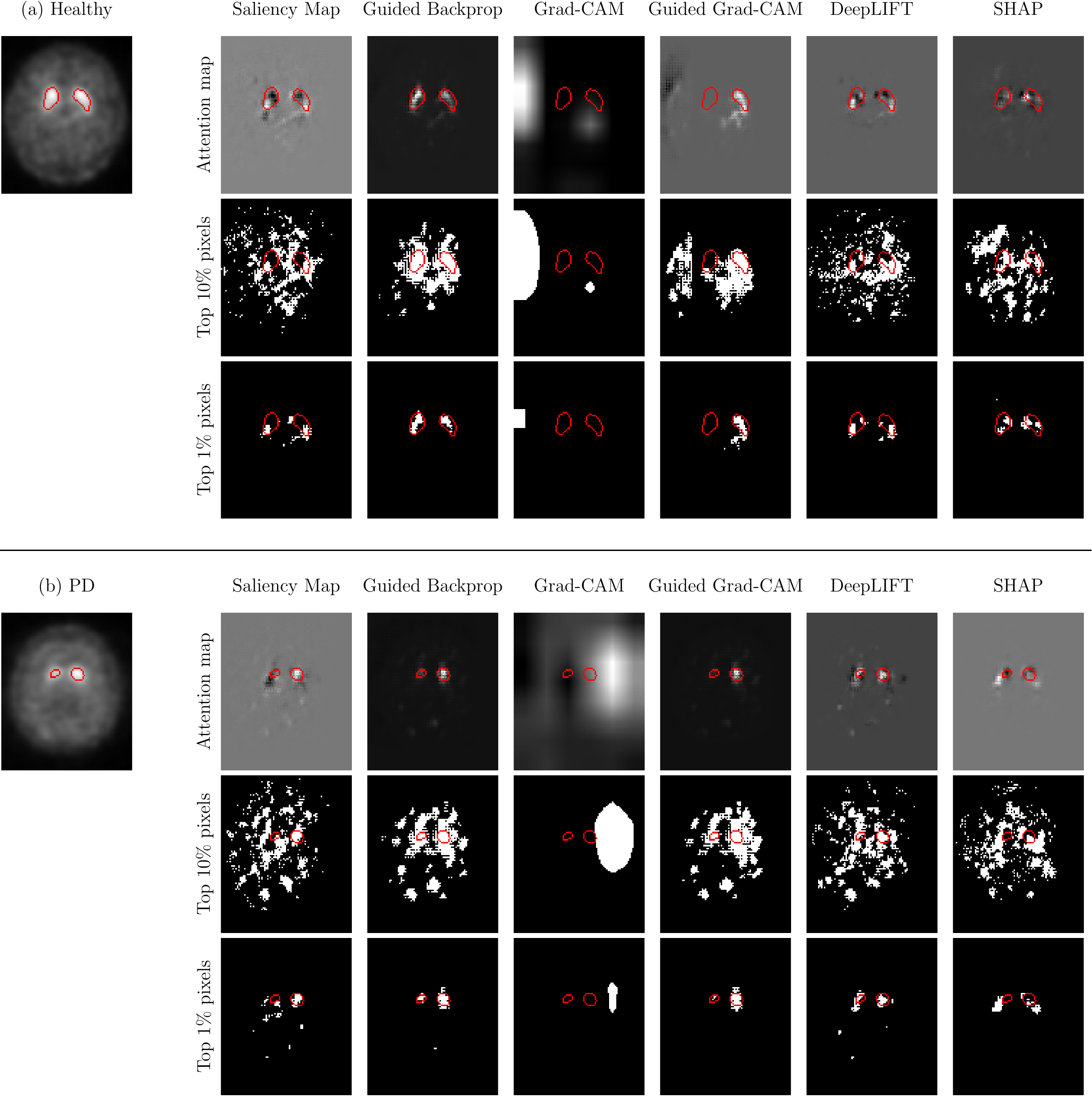}
\caption{An example of slice averaging SPECT image (left figure) and the attention map (right table) from Deep PD Net model for (a) NC and (b) PD. The red line is the segmented line generated from the mean threshold reported in Ref. \citenum{prashanth17}. The first row of the right table shows the original map. The second and the third row shows the binary map generated from the top 10\% of contribution score and the top 1\% contribution score.}
\label{fig:heatmap}
\end{figure*}

\subsection{Interpretation performance Comparison: Example Scenario}
Since there are various model interpretation methods available, it is not possible to decisively express which method is the best for SPECT image classification without actual comparison. Therefore, we demonstrate the interpretation performance of the six well-known interpretation methods mentioned above in this tutorial.

To evaluate the interpretation performance, we generated a ground truth image by segmenting the striatal nuclei. This ground truth image is compared with the attention map from the interpretation methods. The segmented striatal nuclei are created based on a previous study \cite{prashanth17}. The slices from 35th to 48th of the SPECT image, which cover the striatal nuclei, are selected. Then, each slice is normalized to the range from 0 to 1, and a slice averaging image is constructed. This slice averaging image is again normalized to [0,1]. After that, a threshold that determines the segmented area is selected. The mean $\pm$ SD of the thresholds for NC and PD subjects, which the experts selected, were reported in \cite{prashanth17} as 0.63 $\pm$ 0.04 and 0.69 $\pm$ 0.05, respectively. In this tutorial, we select the mean threshold values and use them to find the segmented striatal nuclei of the slice averaging image. The results of the slice averaging SPECT images from NC and PD can be seen in \autoref{fig:heatmap}. The area that is enclosed by the red irregular ellipse represents the segmented area. The segmented area is now used as the ground truth to evaluate the interpretation performance. 

The slice averaging the attention map from the interpretation method was also generated similar to the slice averaging the SPECT images. Examples of grayscale attention maps from the Deep PD Net model are shown in the first row of \autoref{fig:heatmap} (a) and (b) for a NC and a PD subject, respectively. White regions located near or inside the segmented region show the most contributed area in the class prediction.

%%%%%%%%%%%%%%%%%%%%%%%%%%%%%%%%%%%%%%%%%%%%%%%%%%%%%
\begin{table*}
\caption{The results of the mean Dice coefficient using the binary image of the attention map for the top 10\% of contribution scores (upper) and the top 1\% of contribution scores (lower). The bold number refers to the highest Dice coefficient among all the methods.}
\resizebox{\textwidth}{!}{%
	\begin{tabular}{ l l l l l l l l }
Model  & Saliency Map & Guided Backprop & Grad-CAM &	Guided Grad-CAM & DeepLIFT & SHAP &\\
	\hline
PD Net & 17.09 $\pm$ 4.91 & \bf 23.78 $\pm$ 6.02 & 3.27 $\pm$ 8.11 & 22.15 $\pm$ 7.36 & 16.92 $\pm$ 6.75 & 16.62 $\pm$ 6.77 & \\
PD Net + Batch Norm & 12.51 $\pm$ 4.99 & \bf 23.90 $\pm$ 6.76 & 4.49 $\pm$ 7.89 & 20.73 $\pm$ 8.53 & 14.98 $\pm$ 6.04 & 15.50 $\pm$ 6.85 & \\
Deep PD Net & 17.29 $\pm$ 5.00 & \bf 29.72 $\pm$ 8.95 & 3.66 $\pm$ 7.77 & 25.70 $\pm$ 10.15 & 18.11 $\pm$ 6.59 & 15.72 $\pm$ 9.60 & \\
Deep PD Net + Batch Norm & 15.22 $\pm$ 4.36 & \bf 29.38 $\pm$ 9.00 & 2.96 $\pm$ 6.49 & 21.11 $\pm$ 12.16 & 16.99 $\pm$ 5.23 & 16.35 $\pm$ 9.39 & \\
	\hline
\\
\\
Model  & Saliency Map & Guided Backprop & Grad-CAM &	Guided Grad-CAM & DeepLIFT & SHAP &\\
	\hline
PD Net & 38.38 $\pm$ 10.73 & \bf 53.08 $\pm$ 10.42 & 1.45 $\pm$ 5.96 & 49.32 $\pm$ 16.69 & 32.53 $\pm$ 11.53 & 26.73 $\pm$ 11.20 & \\
PD Net + Batch Norm & 22.20 $\pm$ 9.38 & \bf 54.85 $\pm$ 10.12 & 1.85 $\pm$ 6.59 & 47.91 $\pm$ 19.62 & 26.73 $\pm$ 10.27 & 22.63 $\pm$ 11.19 & \\
Deep PD Net & 45.32 $\pm$ 10.02 & \bf 66.07 $\pm$ 12.62 & 1.45 $\pm$ 5.99 & 58.87 $\pm$ 23.86 & 36.96 $\pm$ 11.00 & 25.81 $\pm$ 15.54 & \\
Deep PD Net + Batch Norm & 38.37 $\pm$ 10.22 & \bf 65.56 $\pm$ 12.32 & 0.96 $\pm$ 5.11 & 49.00 $\pm$ 28.71 & 38.71 $\pm$ 10.28 & 28.15 $\pm$ 15.82 & \\
	\hline
	\end{tabular}}
 \label{tbl-overlay}
\end{table*}

\subsection{Evaluation Methods for Interpretable Models}
The pixels that are used to evaluate the interpretation performance need to be selected with another threshold. \cite{shrikumar17} proposed the threshold of which using only 20\% of top values sorted from descending order. In this study, this thresholding technique was used with altering percentages of 10\% and 1\%. Then two binary images can be generated from an attention map. These binary images for different interpretation methods are shown in the second and third row of \autoref{fig:heatmap} (a) and (b), respectively. These figures demonstrate the overlap region between each interpretation method and the segmented area significantly. By considering the figure of the top 10\% pixels as seen in the second row, we can observe that the majority of the pixels are located inside the brain area. On the other hand, the results from using the top 1\% as seen in the third row show that the majority of pixels gather inside the segmented red line area.

\begin{figure}
  \centering
  \includegraphics[width=0.45\textwidth]{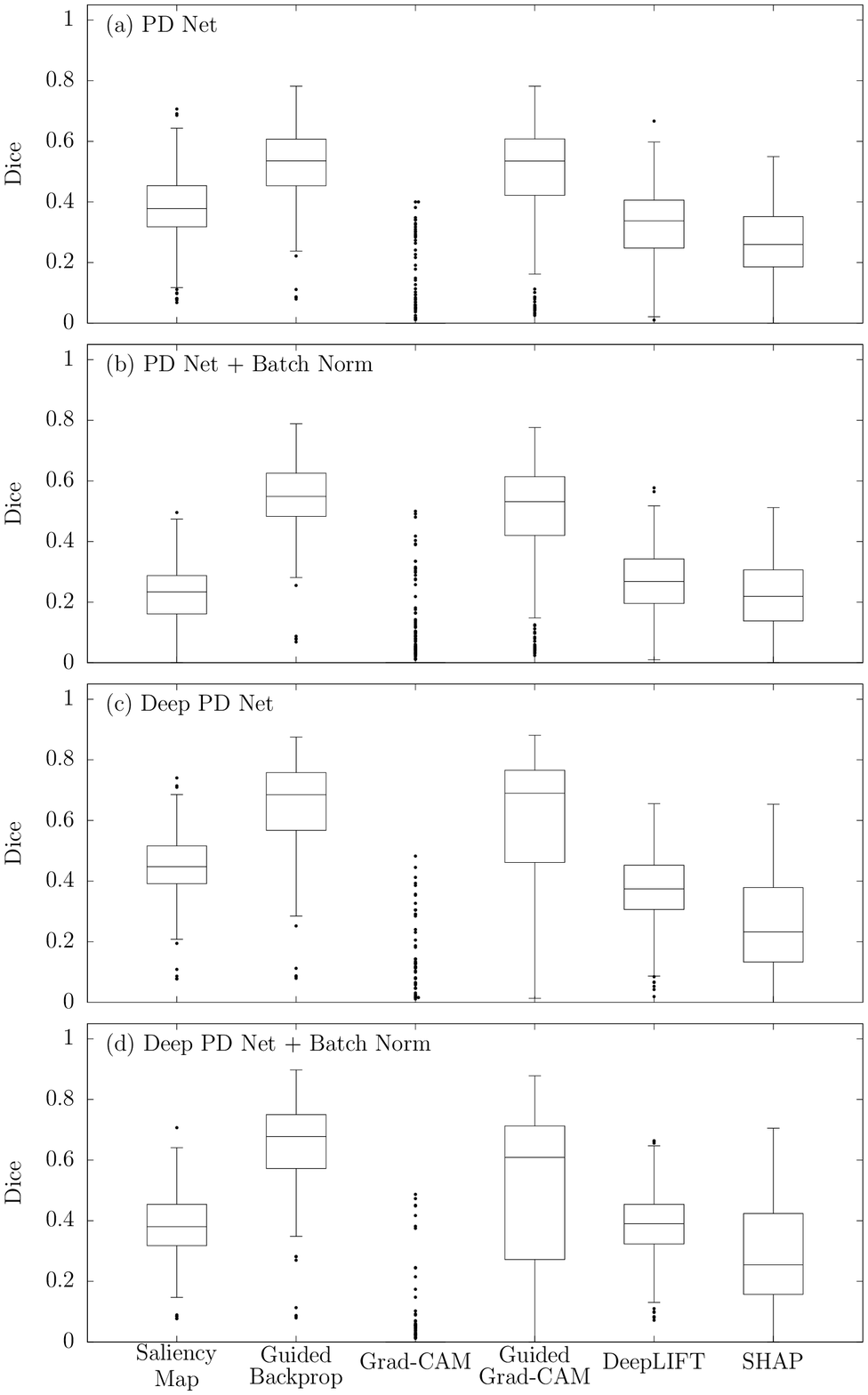}
  \caption{Boxplots of Dice coefficient in different interpretation methods from the top 1\% of contribution score for (a) PD Net (b) PD Net + Batch Norm (c) Deep PD Net and (d) Deep PD Net + Batch Norm. Median is the line that locates inside the box, and black dots represent outliers outside 1.5 times the interquartile range of the upper and lower quartile.}
  \label{fig:dice_box_plot}
\end{figure}

Dice coefficient $D$ is widely used to compare a predicted segmented image $P$ with the ground truth segmented image $G$. It is defined as twice the size of the intersect area between $P$ and $G$ over the sum of the area $P$ and $G$, and can be written as
\begin{equation}
D = \frac{{2\left| {P \cap G} \right|}}{{\left| P \right| + \left| G \right|}}.
\end{equation}
The coefficient exists in the range of $[0,1]$ where $D=1$ indicates identical segmentation. The mean $\pm$ SD of the Dice coefficient is calculated from the test set of all 10-fold. The results are shown in \autoref{tbl-overlay}. The bold value indicates the best result in a given threshold. The upper and lower tables show the results from the top 10\% and top 1\%, respectively. The uses of the top 10\% and 1\% show that guided backpropagation has the highest Dice coefficient, which directly relates to the interpretation performance in providing the information of the location of striatal nuclei. The Dice coefficient's boxplots in \autoref{fig:dice_box_plot} also confirm that guided backpropagation performance dominates other methods. 

Wilcoxon signed-rank test was used to compare the guided backpropagation with the other methods. It revealed significant differences ($p < 0.01$) for the Dice coefficient between guided backpropagation and all other methods. This test was then used to compare the Dice coefficient of guided backpropagation between Deep PD Net and PD Net. We also found the significant difference ($p < 0.01$) of this model. However, the difference between Deep PD Net and Deep PD Net with batch normalization was not significant.

\begin{figure*}
  \centering
  \includegraphics[width=0.9\textwidth]{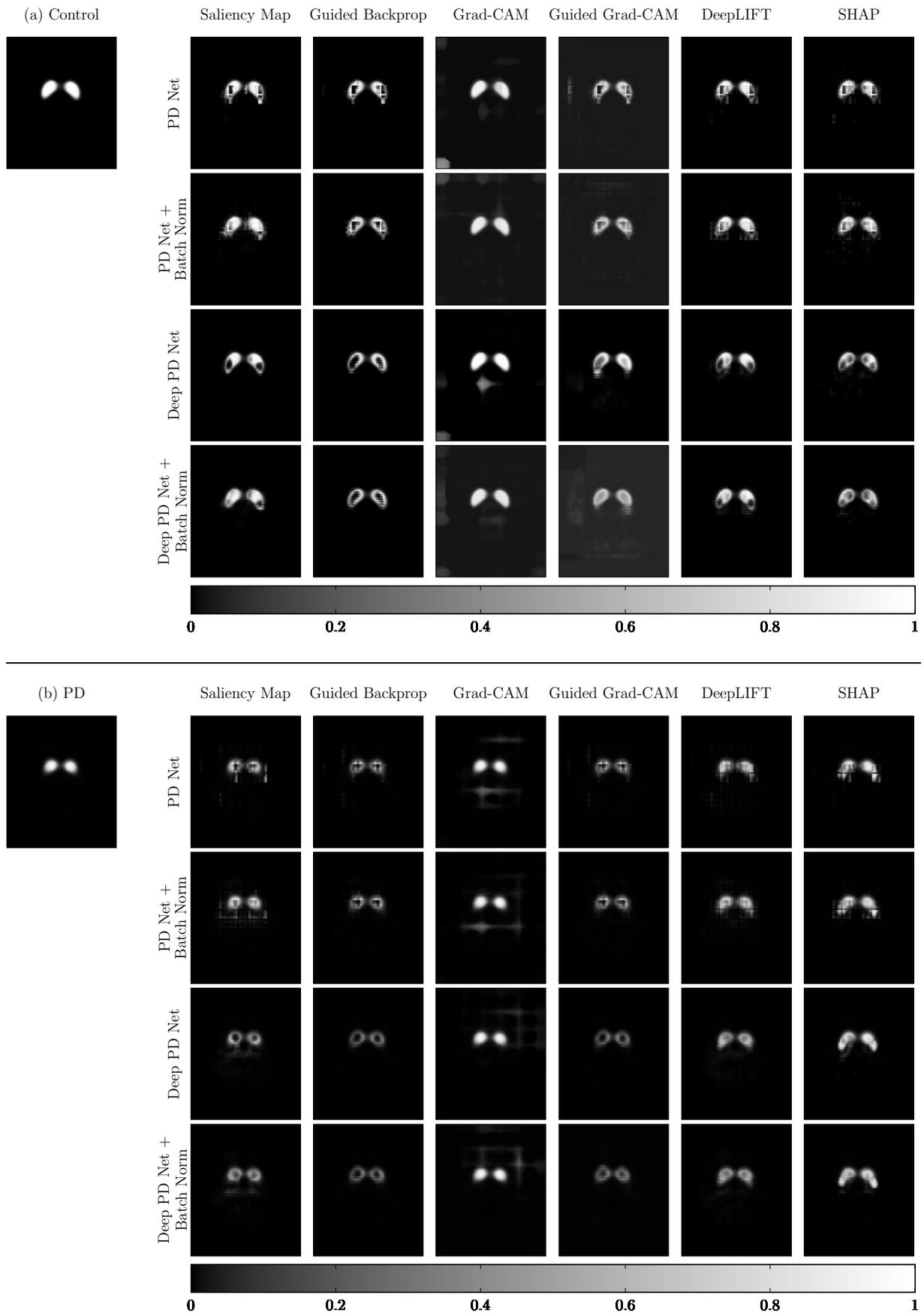}
  \caption{The mean segmented image (left) and mean absolute error plot (right table) for (a) NC group and (b) PD group. The mean absolute error was calculated using the binary image from the top 1\% contribution pixels to compare with the binary image from the segmented image.}
  \label{fig:abs_error}
\end{figure*}

Mean absolute error is used as another measure to evaluate each method's performance as demonstrated in \autoref{fig:abs_error}. The guided backpropagation shows that the error approaches zero inside the striatal nuclei, which can be interpreted as the Deep PD Net directly focuses on the region and gives more credibility to the prediction results.

\begin{figure*}
  \centering
  \includegraphics[width=0.9\textwidth]{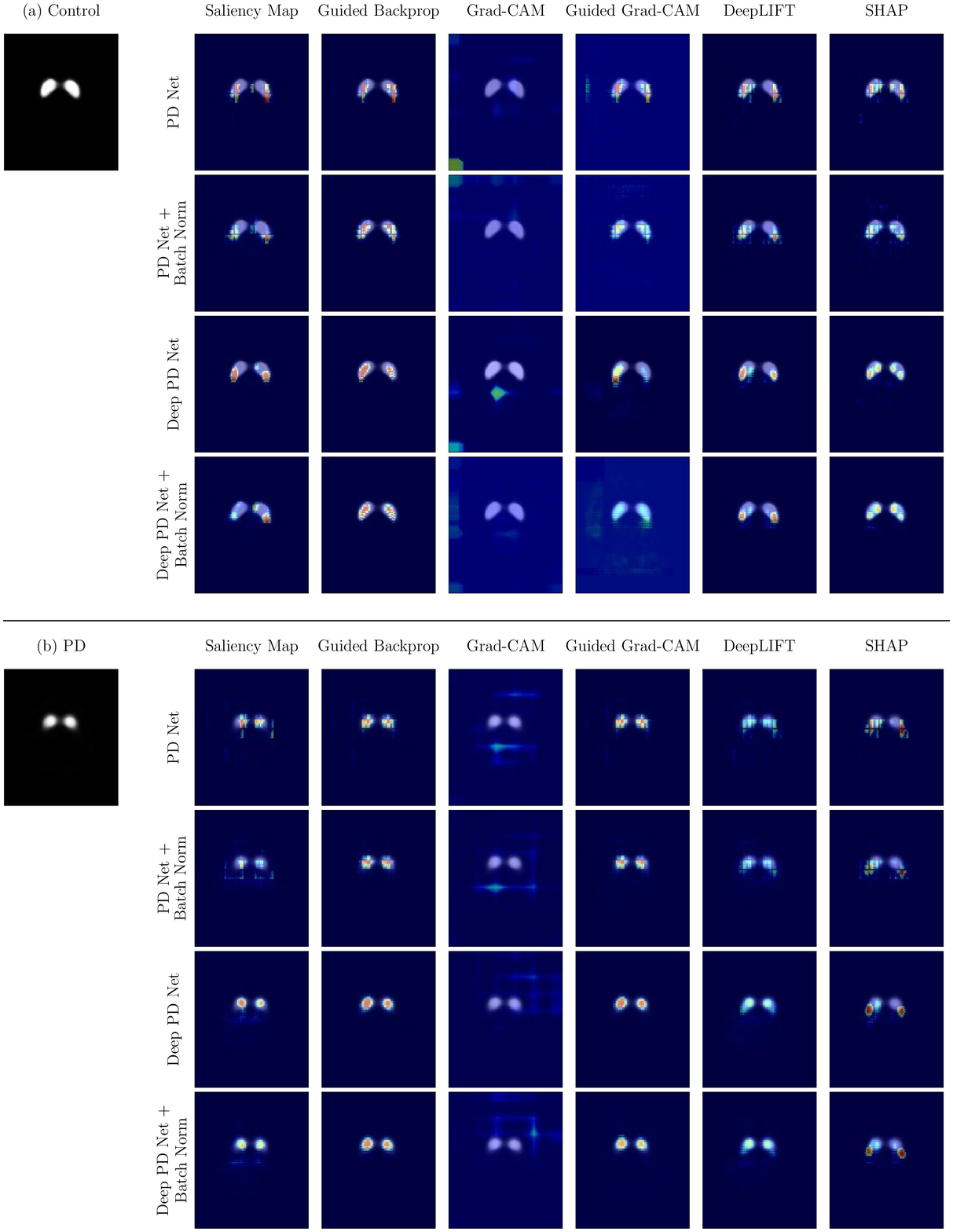}
  \caption{The mean segmented image (left) and mean segmented heatmap (right table) for (a) NC group and (b) PD group. The mean segmented heatmap was generated using the mean of binary images from the top 1\% contribution pixels.}
  \label{fig:mean}
\end{figure*}

We generate a mean segmented image from the binary map of top 1\% pixels and overlay on top of the ground truth segmented image as shown in \autoref{fig:mean}. By examining \autoref{fig:mean}(a), saliency map and DeepLIFT can identify the tail of symmetric comma shape in the control group. On the other hand, from \autoref{fig:mean}(b), SHAP can correctly identify the uptake depletion location of the PD group. 

%%%%%%%%%%%%%%%%%%%%%%%%%%%%%%%%%%%%%%%%%%%%%%%%%%%%

%%%%%%%%%%%%%%%%%%%%%%%%%%%%%%%%%%%%%%%%%%%%%%%%%%%%%
%%%%%%%%%%%%%%%%%%%%%%%%%%%%%%%%%%%%%%%%%%%%%%%%%%

\section{Discussion on Example Scenario}
\label{S:4}
In \autoref{S:2}, we demonstrate the comparison of classification performance between 4 types of DCNN architecture based on the PD Net. DCNN may cause the overfitting of the data \cite{rawat17}. However, Deep PD Net, which attaches more convolution and max-pooling layers to increase the network depth, yields better performance than PD Net without overfitting. The addition of batch normalization to the model shows a minor improvement of the model accuracy, which might be due to the small value of the learning rate set in this study. Also, the input data may not be complex enough compared to the results of the original batch normalization study \cite{ioffe15}. From the clinical details shown in \autoref{tbl-para}, the number of PD subjects is 3 times higher than the number of NC subjects. Due to this extreme class imbalance, we can observe only the increase in the specificity but not the sensitivity.

For the comparison of interpretation performance, the Dice coefficient in \autoref{tbl-overlay} and mean absolute error in \autoref{fig:abs_error} show that guided backpropagation outperforms other methods. Guided backpropagation was first designed to improve the saliency map's quality in feature visualization of the deep learning model \cite{springenberg14}. In this tutorial, it has the best ability to show fine-grained importance. It also gives much less error in the mean absolute error plot compared to other methods. On the other hand, Grad-CAM was the only method that barely focuses on the crucial region. Although Grad-CAM was supposed to perform well in the class-discriminative and localize relevant image regions \cite{selvaraju17}, it barely focuses on fine-grained importance. Another issue of Grad-CAM is that it heavily relies on the resolution of the last feature map. Since PD Net was designed with the last feature map of size $(1\times 1\times 1)$, we need to select the feature map from the upper convolution layer with size $(6 \times 8 \times 6)$ which may reduce the interpretation performance. 

When we plot the mean segmented image from the binary map of the top 1\% pixels, saliency map, DeepLIFT, and SHAP also able to discriminate the difference between classes. Since SHAP is the model-agnostic method, it does not rely on DCNN weights; thus, the result becomes different from the saliency map and DeepLIFT. Even though these three methods' performance is hard to evaluate, SHAP can generate a better quality heatmap at the uptake depletion location, which outperforms other methods in discriminating the difference between PD and NC subjects. This should be consistent with \cite{lundberg17}, which revealed that SHAP gives the best performance among all other methods of showing the class difference between hand-written images of numbers 8 and 3. In conclusion, both guided backpropagation and SHAP are suitable interpretation methods for the architectures in this tutorial. Nevertheless, in other medical image applications, the other interpretation methods might overcome both methods.

Another interesting aspect of the model interpretation is to use the interpreted feedback as an evaluation metric to choose the best model. For example, from \autoref{tbl-acc} and \autoref{fig:roc_curve}, PD Net with batch normalization shows better specificity and AUC value comparing to Deep PD Net, while Deep PD Net shows better accuracy and sensitivity. The performance of the two models is not significantly different and is difficult to evaluate. To this end, the feedback from the interpretation method can provide a decisive answer to the evaluation. From \autoref{tbl-overlay}, when utilizing the guided backpropagation method, Deep PD Net has the highest interpretation performance, which is significantly higher than that of PD Net with batch normalization. This result suggests that Deep PD Net is the best suit for PD recognition among the tutorial's DCNN architecture.

\begin{figure}
  \centering
  \includegraphics[width=0.40\textwidth]{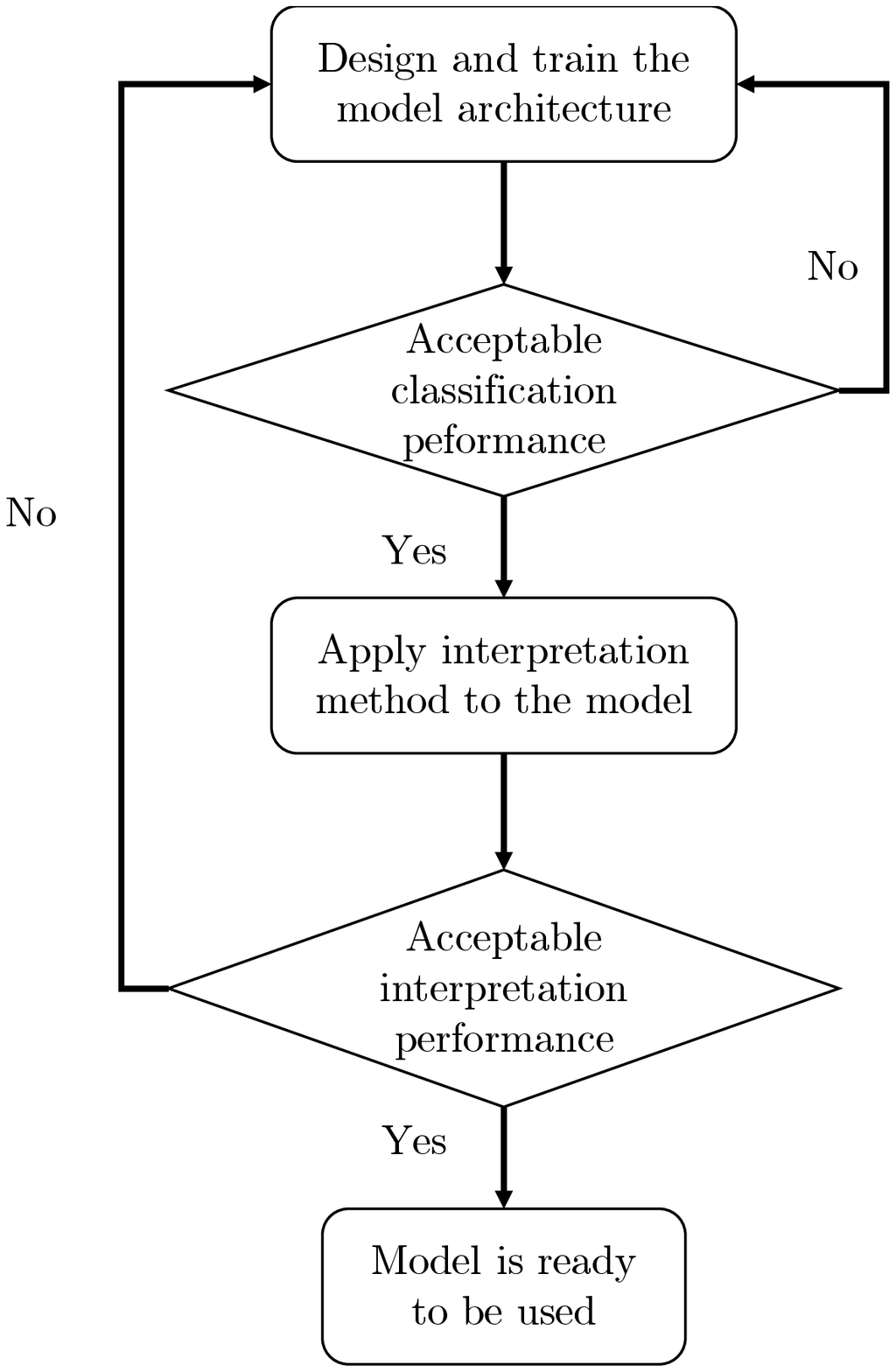}
  \caption{The flow chart for the interpretation method application for assisting in model evaluation.}
  \label{fig:flow_chart}
\end{figure}

We suggest a flow chart for applying the interpreted feedback to assist in model evaluation in \autoref{fig:flow_chart}. In this tutorial, if we follow the flow chart, the Deep PD Net model should be suggested to be used for PPMI data. This flow chart and the concept of model interpretation methods can be utilized not only in medical image application but also in other DCNN application where the credibility of DCNN need to be verified as well, for example, DCNN applications in biomedical \cite{esteva17, lee19} or bioinformatics \cite{le17}.

%%%%%%%%%%%%%%%%%%%%%%%%%%%%%%%%%%%%%%%%%%%%%%%%%%%%%
%%%%%%%%%%%%%%%%%%%%%%%%%%%%%%%%%%%%%%%%%%%%%%%%%%%%%

\section{Conclusions}
\label{S:5}
The purpose of this tutorial is to demonstrate the procedure for selecting the most reliable interpretation method for SPECT image PD recognition model. To this end, we introduce the traditional and DCNN approach for PD diagnosis and give an example scenario with four DCNN models. Then, we introduce six well-known interpretation methods and exhibit these six methods' interpretation performance on those four DCNN models mentioned above. We propose evaluation methods for measuring the interpretation performance using the Dice coefficient and mean segmented binary image overlay on top of ground truth segmented image. {\RED The evaluation demonstrates that the guided backpropagation and SHAP interpretation methods are both suitable for PD recognition methods in different aspects. Guided backpropagation shows the highest Dice coefficient and lowest mean square error, while SHAP provides the best quality heatmap at the uptake depletion location.}  Finally, we discuss about utilizing interpreted feedback for deciding the most suitable model for the intended task. The interpretation and evaluation methods displayed in this tutorial can be applied for other tasks aside from PD recognition, and contribute to sensor data processing in an AI Era.

%\appendix
%\section{Training parameters of DCNN}
%\label{A:1}
%All the DCNN models were implemented with Keras \cite{chollet15}, an open-source deep learning library written in Python and running on top of Tensorflow \cite{abadi17}. The models were trained for 30 epochs using Stochastic Gradient Descent. The momentum parameter was set to 0.9. The learning rate was initially $1 \times 10^{-4}$ and logarithmically decreased to have $1 \times 10^{-6}$ at the final epoch. Additionally, weight parameters in the model were initiated with a Glorot initialization \cite{glorot10}. The loss function also is weighted for class imbalance during the training. These training parameters are the same with \cite{choi17} and every model uses the same parameters for a fair comparison.

\bibliography{bib_ieee}

\begin{thebibliography}{10}
\providecommand{\url}[1]{#1}
\csname url@samestyle\endcsname
\providecommand{\newblock}{\relax}
\providecommand{\bibinfo}[2]{#2}
\providecommand{\BIBentrySTDinterwordspacing}{\spaceskip=0pt\relax}
\providecommand{\BIBentryALTinterwordstretchfactor}{4}
\providecommand{\BIBentryALTinterwordspacing}{\spaceskip=\fontdimen2\font plus
\BIBentryALTinterwordstretchfactor\fontdimen3\font minus
  \fontdimen4\font\relax}
\providecommand{\BIBforeignlanguage}[2]{{%
\expandafter\ifx\csname l@#1\endcsname\relax
\typeout{** WARNING: IEEEtran.bst: No hyphenation pattern has been}%
\typeout{** loaded for the language `#1'. Using the pattern for}%
\typeout{** the default language instead.}%
\else
\language=\csname l@#1\endcsname
\fi
#2}}
\providecommand{\BIBdecl}{\relax}
\BIBdecl

\bibitem{obeso10}
J.~A. Obeso, M.~C. Rodriguez-Oroz, C.~G. Goetz, C.~Marin, J.~H. Kordower,
  M.~Rodriguez, E.~C. Hirsch, M.~Farrer, A.~H.~V. Schapira, and G.~Halliday,
  ``Missing pieces in the {Parkinson's} disease puzzle,'' \emph{Nature
  Medicine}, vol.~16, pp. 653--661, 2010.

\bibitem{chaudhuri09}
K.~R. Chaudhuri and A.~H. Schapira, ``Non-motor symptoms of {Parkinson's}
  disease: dopaminergic pathophysiology and treatment,'' \emph{The Lancet
  Neurology}, vol.~8, no.~5, pp. 464 -- 474, 2009.

\bibitem{djang12}
D.~S. Djang, M.~J. Janssen, N.~Bohnen, J.~Booij, T.~A. Henderson, K.~Herholz,
  S.~Minoshima, C.~C. Rowe, O.~Sabri, J.~Seibyl, B.~N. Van~Berckel, and
  M.~Wanner, ``{SNM} practice guideline for dopamine transporter imaging with
  {123I-Ioflupane} {SPECT} 1.0,'' \emph{Journal of Nuclear Medicine}, vol.~53,
  no.~1, pp. 154--163, 2012.

\bibitem{badiavas11}
K.~Badiavas, E.~Molyvda, I.~Iakovou, M.~Tsolaki, K.~Psarrakos, and N.~Karatzas,
  ``{SPECT} imaging evaluation in movement disorders: far beyond visual
  assessment,'' \emph{European Journal of Nuclear Medicine and Molecular
  Imaging}, vol.~38, no.~4, pp. 764--773, 2011.

\bibitem{taylor17}
J.~C. Taylor and J.~W. Fenner, ``Comparison of machine learning and
  semi-quantification algorithms for {(I123)FP-CIT} classification: the
  beginning of the end for semi-quantification?'' \emph{EJNMMI Physics},
  vol.~4, no.~1, p.~29, 2017.

\bibitem{lecun15}
Y.~LeCun, Y.~Bengio, and G.~Hinton, ``Deep learning,'' \emph{Nature}, vol. 521,
  pp. 436--444, 2015.

\bibitem{goodfellow16}
I.~Goodfellow, Y.~Bengio, and A.~Courville, \emph{Deep Learning}.\hskip 1em
  plus 0.5em minus 0.4em\relax Cambridge: MIT Press, 2016.

\bibitem{hubel62}
D.~H. Hubel and T.~N. Wiesel, ``Receptive fields, binocular interaction and
  functional architecture in the cat's visual cortex,'' \emph{The Journal of
  Physiology}, vol. 160, no.~1, pp. 106--154, 1962.

\bibitem{cadieu14}
C.~F. Cadieu, H.~Hong, D.~L.~K. Yamins, N.~Pinto, D.~Ardila, E.~A. Solomon,
  N.~J. Majaj, and J.~J. DiCarlo, ``Deep neural networks rival the
  representation of primate {IT} cortex for core visual object recognition,''
  \emph{PLOS Computational Biology}, vol.~10, no.~12, pp. 1--18, 12 2014.

\bibitem{litjens17}
G.~Litjens, T.~Kooi, B.~E. Bejnordi, A.~A.~A. Setio, F.~Ciompi, M.~Ghafoorian,
  J.~A. van~der Laak, B.~van Ginneken, and C.~I. S{\'{a}}nchez, ``A survey on
  deep learning in medical image analysis,'' \emph{Medical Image Analysis},
  vol.~42, pp. 60 -- 88, 2017.

\bibitem{duncan20}
J.~S. {Duncan}, M.~F. {Insana}, and N.~{Ayache}, ``Biomedical imaging and
  analysis in the age of big data and deep learning [scanning the issue],''
  \emph{Proceedings of the IEEE}, vol. 108, no.~1, pp. 3--10, Jan 2020.

\bibitem{martinez18}
F.~J. Martinez-Murcia, J.~M. G{\'{o}}rriz, J.~Ram{\'i}rez, and A.~Ortiz,
  ``Convolutional neural networks for neuroimaging in {Parkinson's} disease: Is
  preprocessing needed?'' \emph{International Journal of Neural Systems},
  vol.~28, no.~10, p. 1850035, 2018.

\bibitem{ras18}
G.~Ras, M.~van Gerven, and P.~Haselager, \emph{Explanation Methods in Deep
  Learning: Users, Values, Concerns and Challenges}.\hskip 1em plus 0.5em minus
  0.4em\relax Cham: Springer International Publishing, 2018, pp. 19--36.

\bibitem{esteva17}
A.~Esteva, B.~Kuprel, R.~A. Novoa, J.~Ko, S.~M. Swetter, H.~M. Blau, and
  S.~Thrun, ``Dermatologist-level classification of skin cancer with deep
  neural networks,'' \emph{Nature}, vol. 542, pp. 115--118, Jan 2017.

\bibitem{lee19}
H.~Lee, S.~Yune, M.~Mansouri, M.~Kim, S.~H. Tajmir, C.~E. Guerrier, S.~A.
  Ebert, S.~R. Pomerantz, J.~M. Romero, S.~Kamalian, R.~G. Gonzalez, M.~H. Lev,
  and S.~Do, ``An explainable deep-learning algorithm for the detection of
  acute intracranial haemorrhage from small datasets,'' \emph{Nature Biomedical
  Engineering}, vol.~3, no.~3, pp. 173--182, 2019.

\bibitem{machine_lrp}
J.~{Grezmak}, J.~{Zhang}, P.~{Wang}, K.~A. {Loparo}, and R.~X. {Gao},
  ``Interpretable convolutional neural network through layer-wise relevance
  propagation for machine fault diagnosis,'' \emph{IEEE Sensors Journal},
  vol.~20, no.~6, pp. 3172--3181, 2020.

\bibitem{gait_lrp}
A.~S. {Alharthi}, A.~J. {Casson}, and K.~B. {Ozanyan}, ``Gait spatiotemporal
  signal analysis for parkinson’s disease detection and severity rating,''
  \emph{IEEE Sensors Journal}, vol.~21, no.~2, pp. 1838--1848, 2021.

\bibitem{9115809}
S.~{Lee}, J.~{Kim}, S.~W. {Park}, S.~M. {Jin}, and S.~M. {Park}, ``Toward a
  fully automated artificial pancreas system using a bioinspired reinforcement
  learning design: In silico validation,'' \emph{IEEE Journal of Biomedical and
  Health Informatics}, vol.~25, no.~2, pp. 536--546, 2021.

\bibitem{palumbo10}
B.~Palumbo, M.~L. Fravolini, S.~Nuvoli, A.~Spanu, K.~S. Paulus, O.~Schillaci,
  and G.~Madeddu, ``Comparison of two neural network classifiers in the
  differential diagnosis of essential tremor and {Parkinson's} disease by
  {(123)I-FP-CIT} brain {SPECT},'' \emph{European Journal of Nuclear Medicine
  and Molecular Imaging}, vol.~37, no.~11, pp. 2146--2153, Nov 2010.

\bibitem{palumbo14}
B.~Palumbo, M.~L. Fravolini, T.~Buresta, F.~Pompili, N.~Forini, P.~Nigro,
  P.~Calabresi, and N.~Tambasco, ``Diagnostic accuracy of parkinson disease by
  support vector machine {(SVM)} analysis of {(123)I-FP-CIT} brain {SPECT}
  data: Implications of putaminal findings and age,'' \emph{Medicine}, vol.~93,
  no.~27, p. e228, Dec 2014.

\bibitem{prashanth14}
R.~Prashanth, S.~D. Roy, P.~K. Mandal, and S.~Ghosh, ``Automatic classification
  and prediction models for early {Parkinson's} disease diagnosis from {SPECT}
  imaging,'' \emph{Expert Systems with Applications}, vol.~41, no.~7, pp. 3333
  -- 3342, 2014.

\bibitem{prashanth17}
R.~Prashanth, S.~D. Roy, P.~K. Mandal, and S.~Ghosh, ``High-accuracy
  classification of {Parkinson's} disease through shape analysis and surface
  fitting in 123i-ioflupane {SPECT} imaging,'' \emph{IEEE Journal of Biomedical
  and Health Informatics}, vol.~21, no.~3, pp. 794--802, May 2017.

\bibitem{augimeri16}
A.~Augimeri, A.~Cherubini, G.~L. Cascini, D.~Galea, M.~E. Caligiuri,
  G.~Barbagallo, G.~Arabia, and A.~Quattrone, ``{CADA}---computer-aided
  {DaTSCAN} analysis,'' \emph{EJNMMI Physics}, vol.~3, no.~1, p.~4, Feb 2016.

\bibitem{martinez13}
F.~J. Mart{\'i}nez-Murcia, J.~M. G{\'o}rriz, J.~Ram{\'i}rez, I.~A. Ill{\'a}n,
  and C.~G. Puntonet, ``Texture features based detection of {Parkinson's}
  disease on datscan images,'' in \emph{Natural and Artificial Computation in
  Engineering and Medical Applications}, J.~M. Ferr{\'a}ndez~Vicente, J.~R.
  {\'A}lvarez~S{\'a}nchez, F.~de~la Paz~L{\'o}pez, and F.~J. Toledo~Moreo,
  Eds.\hskip 1em plus 0.5em minus 0.4em\relax Berlin, Heidelberg: Springer
  Berlin Heidelberg, 2013, pp. 266--277.

\bibitem{towey11}
D.~J. Towey, P.~G. Bain, and K.~S. Nijran, ``Automatic classification of
  {I-123-FP-CIT (DaTSCAN) SPECT} images,'' \emph{Nucl Med Commun}, vol.~32,
  2011.

\bibitem{martinez14}
F.~Mart{\'i}nez-Murcia, J.~G{\'{o}}rriz, J.~Ram{\'i}rez, I.~Ill{\'{a}}n, and
  A.~Ortiz, ``Automatic detection of parkinsonism using significance measures
  and component analysis in datscan imaging,'' \emph{Neurocomputing}, vol. 126,
  pp. 58 -- 70, 2014.

\bibitem{segovia12}
F.~Segovia, J.~M. G{\'{o}}rriz, J.~Ram{\'i}rez, I.~{\'{A}}lvarez, J.~M.
  Jiménez-Hoyuela, and S.~J. Ortega, ``Improved parkinsonism diagnosis using a
  partial least squares based approach,'' \emph{Medical Physics}, vol.~39,
  no.~7, pp. 4395--4403, 2012.

\bibitem{rojas13}
A.~Rojas, J.~G{\'{o}}rriz, J.~Ram{\'i}rez, I.~Ill{\'{a}}n,
  F.~Mart{\'i}nez-Murcia, A.~Ortiz, M.~G. R{\'i}o, and M.~Moreno-Caballero,
  ``Application of empirical mode decomposition (emd) on datscan {SPECT} images
  to explore parkinson disease,'' \emph{Expert Systems with Applications},
  vol.~40, no.~7, pp. 2756 -- 2766, 2013.

\bibitem{shiiba20}
T.~Shiiba, Y.~Arimura, M.~Nagano, T.~Takahashi, and A.~Takaki, ``Improvement of
  classification performance of {Parkinson's} disease using shape features for
  machine learning on dopamine transporter single photon emission computed
  tomography,'' \emph{PLOS ONE}, vol.~15, no.~1, pp. 1--12, 01 2020.

\bibitem{illan12}
I.~A. Ill{\'{a}}n, J.~M. G{\'{o}}rriz, J.~Ram{\'{i}}rez, F.~Segovia, J.~M.
  Jim{\'{e}}nez-Hoyuela, and S.~J. Ortega~Lozano, ``Automatic assistance to
  {Parkinson's} disease diagnosis in datscan {SPECT} imaging,'' \emph{Medical
  Physics}, vol.~39, no.~10, pp. 5971--5980, 2012.

\bibitem{oliveira15}
F.~P.~M. Oliveira and M.~Castelo-Branco, ``Computer-aided diagnosis of
  {Parkinson's} disease based on {[123 I]FP-CIT} {SPECT} binding potential
  images, using the voxels-as-features approach and support vector machines,''
  \emph{Journal of Neural Engineering}, vol.~12, no.~2, p. 026008, 2015.

\bibitem{tagare17}
H.~D. Tagare, C.~DeLorenzo, S.~Chelikani, L.~Saperstein, and R.~K. Fulbright,
  ``Voxel-based logistic analysis of {PPMI} control and {Parkinson's} disease
  datscans,'' \emph{NeuroImage}, vol. 152, pp. 299 -- 311, 2017.

\bibitem{zhang17}
Y.~C. Zhang and A.~C. Kagen, ``Machine learning interface for medical image
  analysis,'' \emph{Journal of Digital Imaging}, vol.~30, no.~5, pp. 615--621,
  Oct 2017.

\bibitem{sbr13}
G.~Wisniewski, J.~Seibyl, and K.~Marek, ``{DatScan} {SPECT} image processing
  methods for calculation of striatal binding ratio {(SBR)},'' Institute for
  Neurodegenerative Disorders (IND), Tech. Rep., 2013.

\bibitem{oliveira2015computer}
F.~P. Oliveira and M.~Castelo-Branco, ``Computer-aided diagnosis of
  parkinson’s disease based on [123i] fp-cit spect binding potential images,
  using the voxels-as-features approach and support vector machines,''
  \emph{Journal of neural engineering}, vol.~12, no.~2, p. 026008, 2015.

\bibitem{oliveira2018extraction}
F.~P. Oliveira, D.~B. Faria, D.~C. Costa, M.~Castelo-Branco, and J.~M.~R.
  Tavares, ``Extraction, selection and comparison of features for an effective
  automated computer-aided diagnosis of parkinson’s disease based on [123 i]
  fp-cit spect images,'' \emph{European journal of nuclear medicine and
  molecular imaging}, vol.~45, no.~6, pp. 1052--1062, 2018.

\bibitem{martinez2017}
F.~J. Martinez-Murcia, A.~Ortiz, J.~M. G{\'o}rriz, J.~Ram{\'i}rez, F.~Segovia,
  D.~Salas-Gonzalez, D.~Castillo-Barnes, and I.~A. Ill{\'a}n, ``A 3d
  convolutional neural network approach for the diagnosis of parkinson's
  disease,'' in \emph{Natural and Artificial Computation for Biomedicine and
  Neuroscience}, J.~M. Ferr{\'a}ndez~Vicente, J.~R. {\'A}lvarez-S{\'a}nchez,
  F.~de~la Paz~L{\'o}pez, J.~Toledo~Moreo, and H.~Adeli, Eds.\hskip 1em plus
  0.5em minus 0.4em\relax Cham: Springer International Publishing, 2017, pp.
  324--333.

\bibitem{choi17}
H.~Choi, S.~Ha, H.~J. Im, S.~H. Paek, and D.~S. Lee, ``Refining diagnosis of
  {Parkinson's} disease with deep learning-based interpretation of dopamine
  transporter imaging,'' \emph{NeuroImage: Clinical}, vol.~16, pp. 586 -- 594,
  2017.

\bibitem{wenzel2019automatic}
M.~Wenzel, F.~Milletari, J.~Kr{\"u}ger, C.~Lange, M.~Schenk, I.~Apostolova,
  S.~Klutmann, M.~Ehrenburg, and R.~Buchert, ``Automatic classification of
  dopamine transporter spect: deep convolutional neural networks can be trained
  to be robust with respect to variable image characteristics,'' \emph{European
  journal of nuclear medicine and molecular imaging}, vol.~46, no.~13, pp.
  2800--2811, 2019.

\bibitem{ortiz2019parkinson}
A.~Ortiz, J.~Munilla, M.~Mart{\'\i}nez-Iba{\~n}ez, J.~M. G{\'o}rriz,
  J.~Ram{\'\i}rez, and D.~Salas-Gonzalez, ``Parkinson's disease detection using
  isosurfaces-based features and convolutional neural networks,''
  \emph{Frontiers in neuroinformatics}, vol.~13, p.~48, 2019.

\bibitem{mohammed2021easy}
F.~Mohammed, X.~He, and Y.~Lin, ``An easy-to-use deep-learning model for highly
  accurate diagnosis of parkinson's disease using spect images,''
  \emph{Computerized Medical Imaging and Graphics}, vol.~87, p. 101810, 2021.

\bibitem{ioffe15}
S.~Ioffe and C.~Szegedy, ``Batch normalization: Accelerating deep network
  training by reducing internal covariate shift,'' \emph{Preprint in
  arXiv:1502.03167}, 2015.

\bibitem{Klyuzhin29}
\BIBentryALTinterwordspacing
I.~Klyuzhin, N.~Shenkov, A.~Rahmim, and V.~Sossi, ``Use of deep convolutional
  neural networks to predict parkinson{\textquoteright}s disease progression
  from datscan spect images,'' \emph{Journal of Nuclear Medicine}, vol.~59, no.
  supplement 1, pp. 29--29, 2018. [Online]. Available:
  \url{https://jnm.snmjournals.org/content/59/supplement_1/29}
\BIBentrySTDinterwordspacing

\bibitem{ortiz19}
\BIBentryALTinterwordspacing
A.~Ortiz, J.~Munilla, M.~Martínez-Ibañez, J.~M. Górriz, J.~Ramírez, and
  D.~Salas-Gonzalez, ``Parkinson's disease detection using isosurfaces-based
  features and convolutional neural networks,'' \emph{Frontiers in
  Neuroinformatics}, vol.~13, p.~48, 2019. [Online]. Available:
  \url{https://www.frontiersin.org/article/10.3389/fninf.2019.00048}
\BIBentrySTDinterwordspacing

\bibitem{ppmi12}
K.~Marek \emph{et~al.}, ``The parkinson progression marker initiative
  ({PPMI}),'' \emph{Progress in Neurobiology}, vol.~95, no.~4, pp. 629 -- 635,
  2011.

\bibitem{chollet15}
\BIBentryALTinterwordspacing
F.~Chollet. (2015) Keras. [Online]. Available: \url{https://keras.io/}
\BIBentrySTDinterwordspacing

\bibitem{abadi17}
M.~Abadi \emph{et~al.}, ``{TensorFlow}: Large-scale machine learning on
  heterogeneous distributed systems,'' \emph{Preprint in arXiv:1603.04467},
  2016.

\bibitem{glorot10}
X.~Glorot and Y.~Bengio, ``Understanding the difficulty of training deep
  feedforward neural networks,'' in \emph{Proceedings of the Thirteenth
  International Conference on Artificial Intelligence and Statistics}, ser.
  Proceedings of Machine Learning Research, Y.~W. Teh and M.~Titterington,
  Eds., vol.~9.\hskip 1em plus 0.5em minus 0.4em\relax Chia Laguna Resort,
  Sardinia, Italy: PMLR, 13--15 May 2010, pp. 249--256.

\bibitem{dietterich98}
T.~G. Dietterich, ``Approximate statistical tests for comparing supervised
  classification learning algorithms,'' \emph{Neural Computation}, vol.~10,
  no.~7, pp. 1895--1923, 1998.

\bibitem{shrikumar17}
A.~Shrikumar, P.~Greenside, and A.~Kundaje, ``Learning important features
  through propagating activation differences,'' in \emph{Proceedings of the
  34th International Conference on Machine Learning}, ser. Proceedings of
  Machine Learning Research, vol.~70.\hskip 1em plus 0.5em minus 0.4em\relax
  International Convention Centre, Sydney, Australia: PMLR, 06--11 Aug 2017,
  pp. 3145--3153.

\bibitem{simonyan13}
K.~Simonyan, A.~Vedaldi, and A.~Zisserman, ``Deep inside convolutional
  networks: Visualising image classification models and saliency maps,''
  \emph{Preprint in arXiv:1312.6034}, 2013.

\bibitem{springenberg14}
J.~T. Springenberg, A.~Dosovitskiy, T.~Brox, and M.~Riedmiller, ``Striving for
  simplicity: The all convolutional net,'' \emph{Preprint in arXiv:1412.6806},
  2014.

\bibitem{zhou16}
B.~Zhou, A.~Khosla, A.~Lapedriza, A.~Oliva, and A.~Torralba, ``Learning deep
  features for discriminative localization,'' in \emph{2016 IEEE Conference on
  Computer Vision and Pattern Recognition (CVPR)}, June 2016, pp. 2921--2929.

\bibitem{selvaraju17}
R.~R. Selvaraju, M.~Cogswell, A.~Das, R.~Vedantam, D.~Parikh, and D.~Batra,
  ``Grad-cam: Visual explanations from deep networks via gradient-based
  localization,'' in \emph{2017 IEEE International Conference on Computer
  Vision (ICCV)}, Oct 2017, pp. 618--626.

\bibitem{lundberg17}
S.~M. Lundberg and S.-I. Lee, ``A unified approach to interpreting model
  predictions,'' in \emph{Advances in Neural Information Processing Systems
  30}, I.~Guyon, U.~V. Luxburg, S.~Bengio, H.~Wallach, R.~Fergus,
  S.~Vishwanathan, and R.~Garnett, Eds.\hskip 1em plus 0.5em minus 0.4em\relax
  Curran Associates, Inc., 2017, pp. 4765--4774.

\bibitem{lundberg16}
S.~Lundberg and S.~Lee, ``An unexpected unity among methods for interpreting
  model predictions,'' \emph{Preprint in arXiv:1611.07478}, 2016.

\bibitem{bach15}
S.~Bach, A.~Binder, G.~Montavon, F.~Klauschen, K.-R. M{\"u}ller, and W.~Samek,
  ``On pixel-wise explanations for non-linear classifier decisions by
  layer-wise relevance propagation,'' \emph{PLOS ONE}, vol.~10, no.~7, pp.
  1--46, 07 2015.

\bibitem{rawat17}
W.~Rawat and Z.~Wang, ``Deep convolutional neural networks for image
  classification: A comprehensive review,'' \emph{Neural Computation}, vol.~29,
  no.~9, pp. 2352--2449, 2017.

\bibitem{le17}
N.-Q.-K. Le, Q.-T. Ho, and Y.-Y. Ou, ``Incorporating deep learning with
  convolutional neural networks and position specific scoring matrices for
  identifying electron transport proteins,'' \emph{Journal of Computational
  Chemistry}, vol.~38, no.~23, pp. 2000--2006, 2017.

\end{thebibliography}

\end{document}